\documentclass[conference, 10pt]{IEEEtran}
\IEEEoverridecommandlockouts
\usepackage{cite}
\usepackage{amsmath,amssymb,amsfonts}
\usepackage{algorithmic}
\usepackage{graphicx}
\usepackage{textcomp}
\usepackage{xcolor}
\usepackage{bm}
\usepackage{braket}
\usepackage{subcaption}
\usepackage[a4paper, total={184mm,239mm}]{geometry}
\usepackage[noindentafter]{titlesec}
\def\BibTeX{{\rm B\kern-.05em{\sc i\kern-.025em b}\kern-.08em
    T\kern-.1667em\lower.7ex\hbox{E}\kern-.125emX}}
\begin{document}

\author{\IEEEauthorblockN{
Xiangyu Ren\IEEEauthorrefmark{1},
Junjie Wan\IEEEauthorrefmark{1}, 
Zhiding Liang\IEEEauthorrefmark{2},
Antonio Barbalace\IEEEauthorrefmark{1}}
\IEEEauthorblockA{
\IEEEauthorrefmark{1}
The University of Edinburgh, Edinburgh, UK\\
\IEEEauthorrefmark{2}
Rensselaer Polytechnic Institute, Troy, NY, USA
}}




\title{Tackling Coherent Noise in Quantum Computing\\ 
via Cross-Layer Compiler Optimization}

\maketitle

\begin{abstract}
Quantum computing hardware is affected by quantum noise that undermine the quality of results of an executed quantum program. 
Amongst other quantum noises, coherent error that caused by parameter drifting and miscalibration, remains critical. While coherent error mitigation has been studied before, studies focused either on gate-level or pulse-level
-- missing cross-level optimization opportunities; 
And most of them only target single-qubit gates 
-- while multi-qubit gates are also used in practice. 

To address above limitations, this work proposes a cross-layer approach for coherent error mitigation that considers program-level, gate-level, and pulse-level compiler optimizations, 
by leveraging the hidden inverse theory, and exploiting the structure inside different quantum programs, 
while also considering multi-qubit gates.
We implemented our approach as compiler optimization passes, and integrated into IBM Qiskit framework. We tested our technique on real quantum computer (IBM-Brisbane), and demonstrated up to $92\%$ fidelity improvements ($45\%$ on average), on several benchmarks.

\end{abstract}


\section{Introduction}
Quantum computing has shown potential in solving many classically intractable problems~\cite{boixo2018characterizing}. 
Among diverse quantum computing hardware, superconducting quantum qubit demonstrates fast operation times, making it a promising path towards quantum advantage~\cite{arute2019quantum}.
However, in current Noisy Intermediate Scale Quantum (NISQ) era, quantum computers may still produce flawed results due to effect of quantum noise during program execution~\cite{preskill2018quantum}.

\textit{Incoherent} error and \textit{coherent} error are two distinct types of quantum noise.
Because of those error sources, quantum error mitigation (QEM) techniques have been developed to suppress the impact of quantum noise. 
For example, probabilistic error cancellation (PEC)~\cite{van2023probabilistic} and zero noise extrapolation (ZNE)~\cite{giurgica2020digital} are two well known QEM protocols that show significant improvement.
Based on their error model assumptions, PEC and ZNE are more suitable for correcting \textit{incoherent} errors~\cite{he2020zero, gutierrez2016errors}.
On the other side, the \textit{coherent} error caused by imperfect classical control and parameter drift, remains critical. 
Even if the coherent noise accounts for a small fraction of total error rate, its quadratically worse impact could be much harmful over the execution of a quantum program~\cite{gokhale2020optimized}.

A variety of techniques are put forward to mitigate \textit{coherent} error, and some of them optimize on the quantum circuit: 
Random compiling technique~\cite{wallman2016noise} runs multiple modified versions of the quantum circuit to cancel out the error; 
Error compensation technique~\cite{endo2018practical,strikis2021learning} adjusts the parameter of gates in quantum circuit, relying on the information of error model. 
Despite such efforts, current techniques for mitigating \textit{coherent} errors are still inadequate.
For example, in the case of circuit-level optimization techniques, the program has to be executed for an exponential number of times -- costing a lot of quantum computing resources. Further, there is a strong dependency on the precision of error model retrieved from machine calibration -- which is not robust to timely drifting of control parameters. 

Apart from circuit-level techniques, there are pulse-level techniques such as optimal pulse control~\cite{werschnik2007quantum, koch2022quantum, caneva2011chopped} and pulse ansatz~\cite{liang2024napa, liang2023spacepulse, egger2023pulse}, aiming to improve the classical control of quantum gates by optimizing the pulse parameters.
However, existing pulse-level optimization works are usually restricted to single-qubit gates, hence neglecting the opportunities arising with two-qubit and multi-qubit gates. 
Moreover, existing works 
didn't consider optimization chances brought by quantum circuit's structure.

\begin{figure}[t]
\vspace{-3mm}
    \centering
    \includegraphics[page=1,width=.47\textwidth]{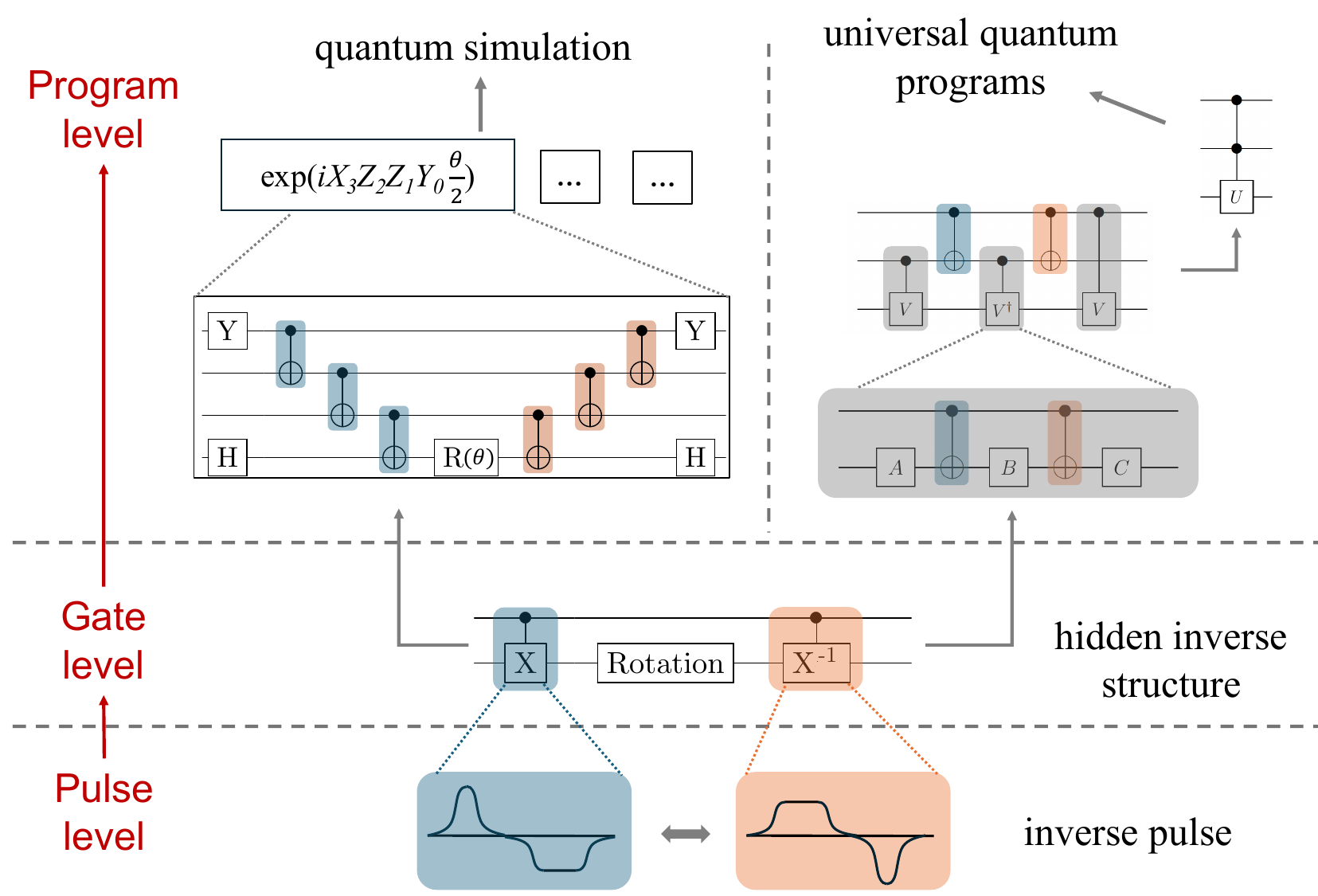}
    \caption{Overview of our cross-layer compiler optimization.} 
    \label{fig:overview}
\vspace{-3mm}
\end{figure}

To address above limitations 
we propose a cross-layer optimization framework (Fig.~\ref{fig:overview}) combining both the advantages of circuit-level and pulse-level methods. 
From the pulse-level to gate-level~\cite{liang2023hybrid}, we find a quantum pulse control pattern that can mitigate coherent noise under specific gate structure (e.g. $CX$ gate pairs), based on the theory of \textit{hidden inverse}~\cite{zhang2022hidden}. From the gate-level to program-level, we exploit the characteristics of different quantum algorithms, developing compiler optimization passes to unveil the specific gate structure hidden in the circuits, thus seizing the opportunities to mitigate coherent noise on such gate structure.

Our key contributions can be summarized as:
\begin{enumerate}
\item We introduce a \textbf{pulse-level} method targeted at a specific \textbf{gate pattern} -- \textit{hidden inverse} to mitigate coherent errors, and validate its feasibility on a real quantum computer.

\item By exploiting the structures of different \textbf{quantum programs}, we develop a compilation method that proactively unveil the \textit{hidden inverses} inside a circuit, creating the chance for coherent noise mitigation based on the pulse-level method in 1).

\item With a combination of 1) and 2),  we develop a \underline{cross-layer} \underline{compiler optimization framework}, and test its effect of mitigating coherent error on a real quantum computer.
\end{enumerate}

In this paper, we focus on the superconducting quantum computers, however our framework is also applicable to other types of quantum hardware such as trapped-ion, just with modification on the control pulse method~\cite{majumder2023characterizing}.

\section{Background}

\subsection{Coherent Error}
Quantum errors are defined as any evolution of the qubits that differs from the ideal intended operation. 
They may be categorized in to different types, e.g., incoherent error, coherent error, state preparation and measurement (SPAM) error, and leakage error.
Among these, coherent error refers to the error that can be represented as a single unitary operator, mapping pure states to pure states. 
It is induced by several sources, such as miscalibration and drift-out-of-calibration in the control system, or crosstalk interactions between neighbouring qubits.

In this paper, we focus on the coherent errors caused by miscalibration and drifting, and our research is orthogonal and complementary to crosstalk suppression techniques~\cite{niu2022effects,niu2024multi}.
From the perspective of quantum circuit execution, such error reflects on the over-rotation of quantum rotation gates and quantum phase misalignment. On the quantum hardware side, quantum engineering try to minimize the negative impact by frequently calibrating the qubits. For example, IBM Quantum cloud platform perform daily calibration on the single-qubit and two-qubits gates, with the job takes 30-90 minutes once. However, the coherent error is still inevitable due to limitation of precision for current quantum control systems. Hence, mitigating coherent error via software techniques is vital for suppressing quantum noise for the NISQ quantum computers.

\begin{figure}[h]
\vspace{-3mm}
    \centering
    \includegraphics[page=1,width=.4\textwidth]{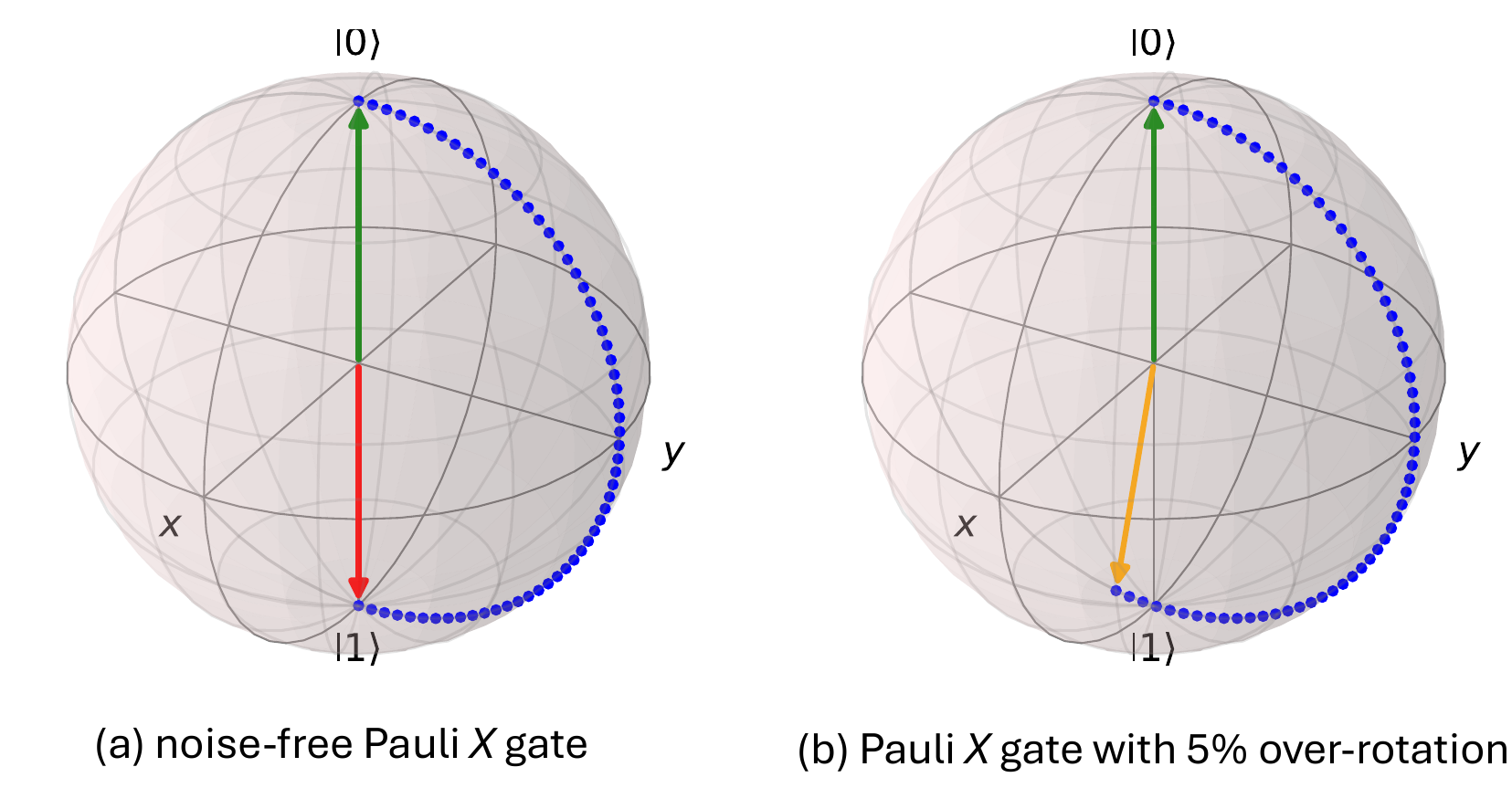}
    \vspace{-3mm}
    \caption{Over-rotation error of a Pauli $X$ applied on state $\ket{0}$.} 
    \label{fig:coherent_noise}
\vspace{-3mm}
\end{figure}

{\textbf{Modeling the coherent error.}}
Based on the understanding of system interaction Hamiltonian, it is often possible to describe coherent errors as additive or multiplicative terms in the control parameters. Taking the rotation-X gate ($R_X(\theta)$) as an example, it can be represented as the Hamiltonian form $exp(-i \frac{\theta}{2} X)$, while $\theta$ is the ideal rotation angle. Given a unwanted rotation error be fraction $\epsilon$ on this gate, it turns out to be $exp(-i \frac{\epsilon\theta}{2} X)$. Fig.~\ref{fig:coherent_noise} gives an illustration of the over-rotation of Pauli $X$ gate on the Bloch sphere.

For the two-qubit gate $CX$, a.k.a. controlled-NOT, it is decomposed into machine supported gates: cross-resonance (CR) gate (a.k.a. $R_{ZX}$ rotation) and other single qubit rotations. Each of these gates induce their over-rotation $\epsilon$, leading to a set of $exp(-i \frac{\epsilon_0\theta}{2} ZX)$, $exp(-i \frac{\epsilon_1\theta}{2} X)$, ..., successively performed on the qubits. Hence, combination of these error makes a complex noise model on the course of quantum circuit.

\subsection{Quantum Pulse Control}
For a compiler of superconducting quantum computers, the pulse-level workflow transforms quantum gates into control pulses applied on hardware. 
Pulses are represented as the waveform of analog signals that applied to quantum channels of different qubits. 
Taking IBM's quantum computer as an example, the \textit{Drive-Channel} is responsible for primary control of each qubit, while \textit{Control-Channel} is only associated with multi-qubit gate operations.
In addition, there are \textit{Acquire-Channel} and \textit{Measure-Channel} activated during qubit measurement and readout operations.

In this paper, we focus on pulses of \textit{Drive-Channel} and \textit{Control-Channel} that is related to quantum gates.
Looking at the pulse scheduling of $CX$ (Fig.~\ref{fig:cxpulse}(a)), the X-axis is scheduling timeline, while on Y-axis there are 
\textit{Drive-Channel} ($D\#$) and \textit{Control-Channel} ($U\#$) waveforms.
Each pulse is associated with a set of parameters: amplitude, duration, angle, width, etc.
These parameters are pre-defined from the hardware calibration process, but users can change 
those for optimization purposes.

\section{Motivation}
\subsection{Gaps in Previous Research}

\noindent\textbf{Circuit-Level Methods.}
\textit{Randomized compiling}~\cite{wallman2016noise} technique is developed based on the Pauli twirling~\cite{magesan2011scalable} method. Random Pauli gates are inserted into original circuit to twirl the coherent errors of quantum gates. Then with executions of different modified circuits, their results are averaged to obtain a noise-mitigated result. Random compiling requires an exponential times of circuit execution, and sometimes it will extend the circuit depth, hence its scalability is not promising.
\textit{Error compensation}~\cite{endo2018practical,strikis2021learning} technique compensate the over-rotation error, by inserting a compensation gate that inverts the noisy gate $R_Z(\epsilon\theta)$ back to ideal gate $R_Z(\theta)$. The compensation gate is decided based on the noise model given from calibration, however it becomes imprecise with the timely drifting of parameters.

\begin{figure*}
    \begin{subfigure}[b]{1\columnwidth}
        \centering
        \includegraphics[width=0.85\columnwidth]{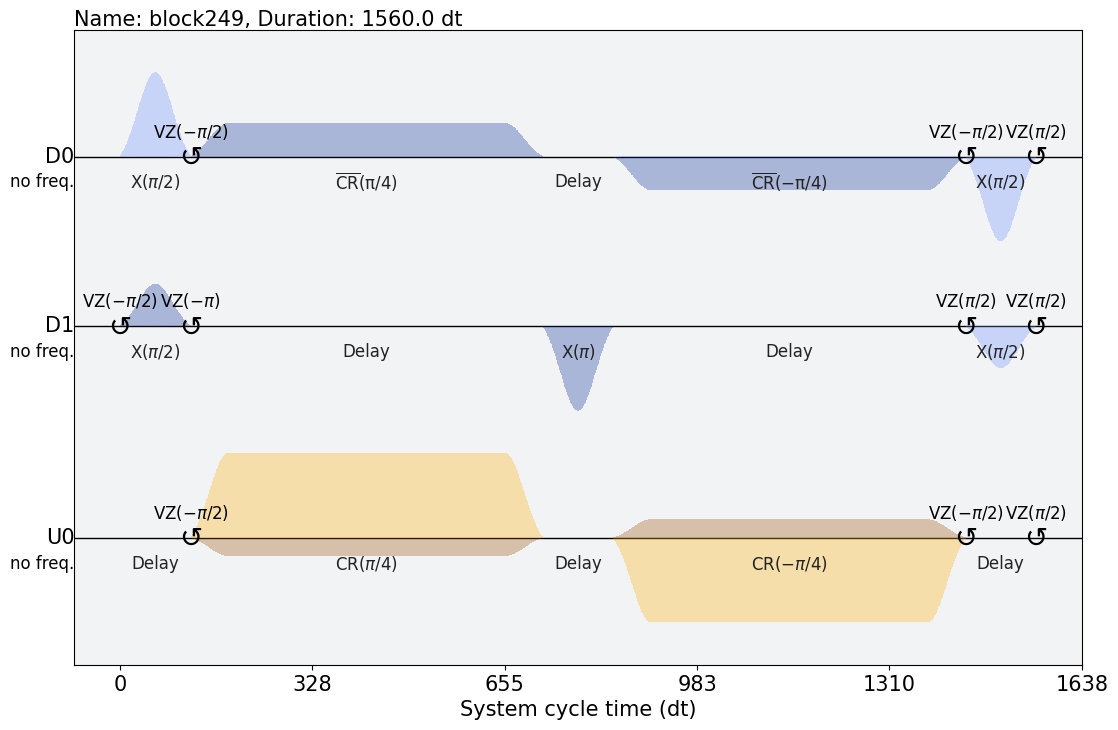}
    \end{subfigure}
    \begin{subfigure}[b]{1\columnwidth}
        \centering
        \includegraphics[width=0.85\columnwidth]{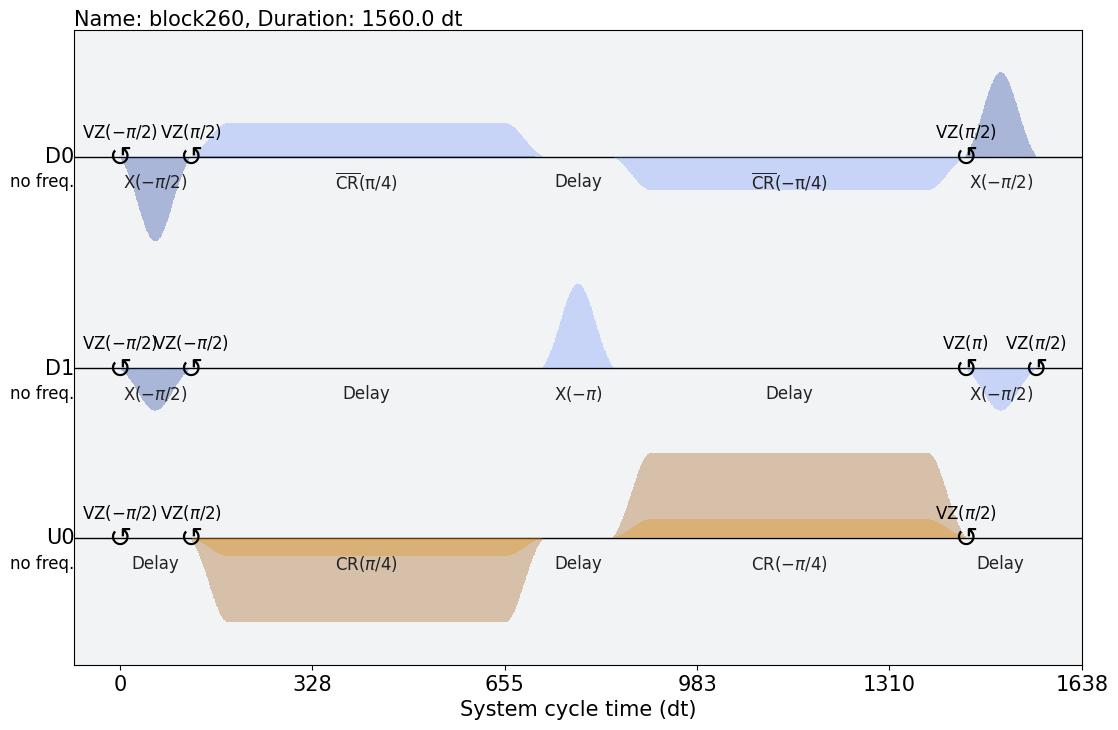}
    \end{subfigure}
    \caption{(Left): CX gate control pulse generated with Qiskit Pulse, with the configuration of IBM Brisbane quantum computer. (Right): Gate control pulse of the $CX^{-1}$ gate, constructed using the parameters in original $CX$. The pulses are not fully symmetric to the original pulses due to phase difference.}
    \label{fig:cxpulse}
    \vspace{-3mm}
\end{figure*}

\noindent\textbf{Pulse-Level Methods.} 
\textit{Quantum optimal control}~\cite{werschnik2007quantum, koch2022quantum, caneva2011chopped} leverages gradient descend to find the best shape and parameters of the control pulse waveform, thus improving fidelity of quantum gates. Optimal control is limited to each single gate, but missing the high-level circuit information that may help to optimize on the execution of quantum program.
\textit{Pulse ansatz}~\cite{liang2024napa, liang2023spacepulse, egger2023pulse} aims to optimize the pulse towards Variational Quantum Algorithm (VQA) programs, by constructing ansatz with pulse parameters that can be directly trained during learning tasks.

\subsection{Opportunities of Cross-Layer Optimization}
To address the limitations of previous research, we aim to design a model that combines both the advantage of circuit-level and pulse-level optimizations. 
From pulse-level to gate-level, we seek the chance to suppress coherent error by transforming the waveform parameters to achieve higher fidelity of quantum gates, hence optimizing each element in a quantum circuit. 
From gate-level to circuit-level, we aim to analyze the hidden structure of different quantum programs. By finding a frequent pattern of the quantum circuit, we are able to utilize circuit structure information to extend our pulse optimization to a higher level.
Moreover, we want to constructing a compiler pipeline that integrates these multiple layer optimization, applying to universal quantum computing programs.

\section{Gate-Pulse Co-optimization with \\ Inverse Pulse}

\subsection{Inverse Pulse}
\noindent\textbf{Coherent noise of two-qubit gates.}
The coherent error of a two-qubit quantum gate has been an unneglectable component in quantum noise, because the implementation of two-qubit gate is more complex than the single-qubit gate. 
According to the noise information of IBM-Brisbane quantum computer, the median error rate of two-qubit gates ($7.7\times10^{-3}$) is approximately 30$\times$ higher than the single-qubit gates ($2.5\times10^{-4}$). 
Among different two-qubit gates, $CX$ gate plays a major part in universal quantum computing programs, stressing the necessity of mitigating the coherent error of $CX$. 
Next, we introduce a gate-pulse co-optimization to tackle coherent error, utilizing the method of \textbf{inverse pulse}.

\noindent\textbf{Introducing the inverse version of $\boldsymbol{CX}$.}
An important characteristic of $CX$ is being self-adjoint, meaning that it is equal to its own inverse: $CX = CX^{-1}$. For a self-adjoint unitary, its inverse version can be constructed by the inverse of its composites. E.g., if $CX = ABC$ then we have $CX^{-1} = C^{-1}B^{-1}A^{-1}$ which implements exactly the same unitary. Thus, to apply a $CX$ gate, it's applicable to use either the standard ($CX$) or the inverse ($CX^{-1}$) version. 

\noindent\textbf{Constructing the inverse pulse of $\boldsymbol{CX}$.}
In Fig.~\ref{fig:cxpulse} we show the pulse of $CX$ gate, which is composed of two $CR$ gate pulses and several single qubit rotation pulses. Base on the self-adjoint characteristic of $CX$, we can construct the pulse of $CX^{-1}$, shown in Fig.~\ref{fig:cxpulse}(b). Here we just utilize the same operations that composes $CX$ pulse, but adjust the pulse amplitudes to its opposite and invert their sequence~\cite{leyton2022quantum}.



\noindent\textbf{Error mitigation within the inverse gate pair.}
Our error mitigation technique is originated from the theoretical work of \textit{hidden inverse}~\cite{zhang2022hidden}. We exploit a common structure in the multi-qubit gates such as control-$R_Z$ ($CR_Z$) and control-$Phase$ ($CPhase$), which are used frequently used in various algorithms. They are compiled into a single-qubit rotation sandwiched by two $CX$ gates at each side, with Fig.~\ref{fig:crzdecomposition}~(a) as an example. By replacing one of the $CX$ with $CX^{-1}$, these two $CX$ gates form a \textit{hidden inverse}. Based on the characteristic of inverse pulse, we suppose that the pair of $CX$ and $CX^{-1}$ share the same sets of error fraction $\epsilon_0, \epsilon_1, ..., \epsilon_n$. It is theoretically proven in \cite{zhang2022hidden} that coherent error within these two gates can be partially cancelled out with each other. By general explanations, in the inverse pulse, over-rotation and phase error are twirled to an opposite direction, which leads to the error mitigation.

\begin{figure}[h!]
\vspace{-3mm}
    \centering
    \includegraphics[page=1,width=.36\textwidth]{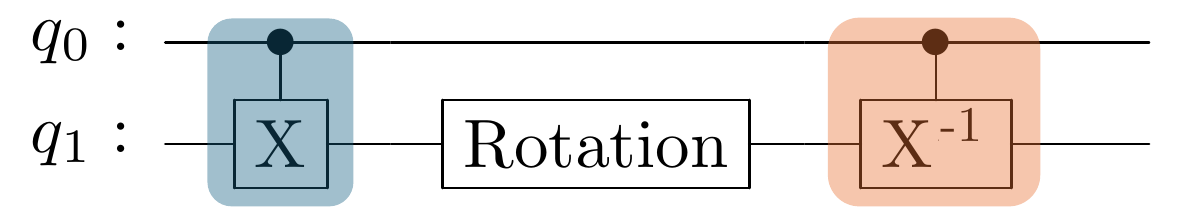}
    \caption{A typical structure of hidden inverse on two qubits: a single-qubit rotation gate, sandwiched by $CX$ and $CX^{-1}$ on each side.} 
    \label{fig:crzdecomposition}
\vspace{-2mm}
\end{figure}

This leads to our gate-pulse co-optimization in quantum circuit: Identifying this structure within the decomposition of various multi-qubit gates, and mitigating the coherent error by replacing $CX$ with $CX^{-1}$ to construct \textit{hidden inverses}. In the following subsection, we verify the effect of hidden inverse on real hardware, by studying a typical case -- the $CR_Z$ gate.

\subsection{A Case Study: Inverse Pulse in $CR_Z$ Gate}
Here we apply the hidden inverse technique to $CR_Z(\theta)$ gate on a superconducting quantum computer, and analysis its effect using a benchmarking circuit.

\noindent\textbf{Decomposition of the $CR_Z$ Gate.}
The details are illustrated in Fig.~\ref{fig:crzdecomposition}. We start with the decomposition of $CR_Z(\theta)$, which leads to two $CX$ gates sandwiching a $R_Z(-\frac{\theta}{2})$ in the middle, forming the structure that can be optimized by applying the \textit{hidden inverse}. Besides, there is another $R_Z(\frac{\theta}{2})$ on the left, which is not included in the structure.

\noindent\textbf{Constructing a inverse $\boldsymbol{CR_Z(\theta)}$ gate.}
Specifically, we change the decomposition of $CR_Z(\theta)$ gate into the form shown in Fig.\ref{fig:crzdecomposition}. 
In the pair of $CX$ gates sandwiching a $R_Z$, we substitute the one on the right with its inverse version $CX^{-1}$, based on their equality ($CX = CX^{-1}$).
The $CX^{-1}$ gate here is implemented via the inverse pulse described in Section.~IV.~a.

\begin{figure}[h!]
    \centering
    \includegraphics[page=1,width=.45\textwidth]{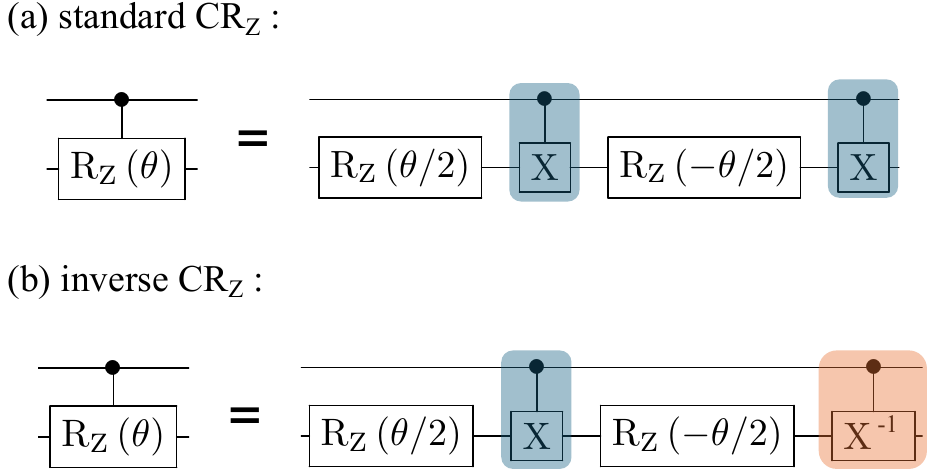}
    \caption{Standard and inverse decomposition of $CR_Z$ gate.} 
    \label{fig:crzdecomposition}
\vspace{-3mm}
\end{figure}

\begin{figure}[h!]
    \centering
    \includegraphics[page=1,width=.48\textwidth]{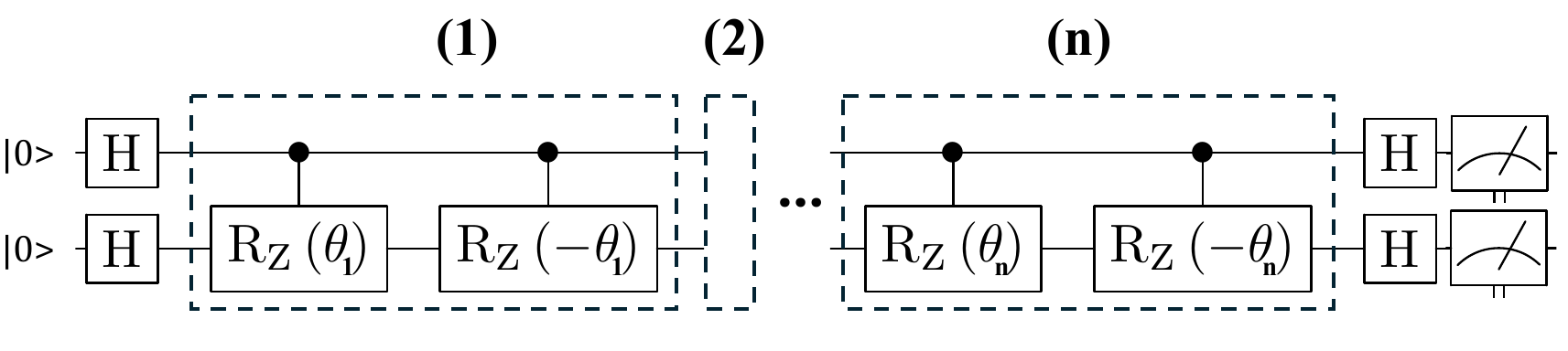}
    \caption{The benchmarking circuit for $CR_Z$ gate.} 
    \label{fig:crzbenchmarkcirc}
\end{figure}

\noindent\textbf{Benchmarking $\boldsymbol{CR_Z(\theta)}$ on real hardware.}
We design the random benchmark circuit shown in Fig.~\ref{fig:crzbenchmarkcirc}, to validate if \textit{hidden inverse} optimization mitigates the coherent error.
The benchmark circuit is inspired from \textit{unitary folding test}~\cite{giurgica2020digital} that scale the noise of specific unitary.
With each dashed line box in Fig.~\ref{fig:crzbenchmarkcirc}, we have a $R_Z(\theta_k)$ and $R_Z(-\theta_k)$, with the rotation angel $\theta_k$ are sampled randomly from $[0,\pi)$. Such box is repeated $n$ times in the circuit, as the length of benchmark circuit. With each box implementing an identity gate ($R_Z(-\theta_k)\cdot R_Z(\theta_k) = I$), the circuit result correspond to initial state under a noise-free assumption.

The random benchmark circuits are run on the IBM-Brisbane quantum computer. 
The initial state of two qubits are set to $\ket{00}$, with Hardamard gates both in the beginning and at the end of the circuit, then finally measure them in Z-axis.
Each rotation angle $\theta$ is sampled uniformly and randomly, and the final result is a average of all circuit instances.
Ideally all the circuit instances could obtain $\ket{00}$ in the result, but noise deviate them from getting the correct answer. Thus, we compare the probability of getting $\ket{00}$ in result, between the standard version and inverse version of $R_Z$.

\begin{figure}[h!]
    \centering
    \includegraphics[page=1,width=.47\textwidth]{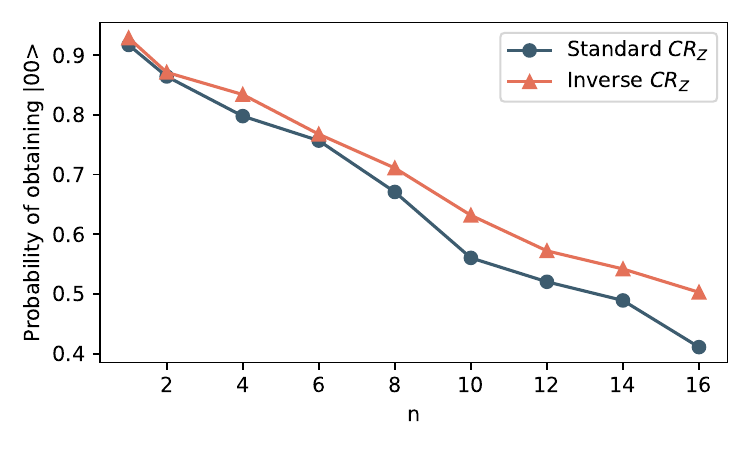}
    \caption{The benchmarking result for CRZ gate. On X-axis, $n$ is the number of repeating of dashed boxes in Fig.~\ref{fig:crzbenchmarkcirc}.} 
    \label{fig:crzbenchmarkresult}
\vspace{-3mm}
\end{figure}

From the data shown in Fig.~\ref{fig:crzbenchmarkresult}, there is a improvement in the probability of getting correct result by using the inverse $CR_Z$, comparing to the standard $CR_Z$. Also, the improvement becomes more significant with the increase on $n$.

\section{Unveiling Hidden Inverse in Circuit \\ via Cross-Layer Optimization}
In the last section we introduce a gate-pulse co-optimization approach based on the \textit{hidden inverse}, and validate its feasibility via random benchmarking. 
however, quantum circuit are very different, and most of the circuits do not just contain $CR_Z$ gates. thus, it's non-trivial to apply the gate-pulse co-optimization to any quantum circuits.
This calls for a systematic approach to identify the structure in a quantum circuit that holds a opportunity to be improved by \textit{hidden inverse}.

In this section, we first focus on a specific quantum algorithm: the quantum simulation kernel~\cite{li2022paulihedral}, and demonstrate how we learn from the program-level information to optimize.
Then, we will discuss the program-gate co-optimization for universal quantum programs. 
Finally, we build up our model for a cross-layer compiler optimization to tackle the coherent noise.

\subsection{Program-Based Optimization}
To demonstrate how program-based information can benefit our compiler optimization model, quantum simulation programs~\cite{li2021software, li2022paulihedral,lao20222qan} are chosen as an example.

\begin{figure}[h!]
    \centering
    \includegraphics[page=1,width=.48\textwidth]{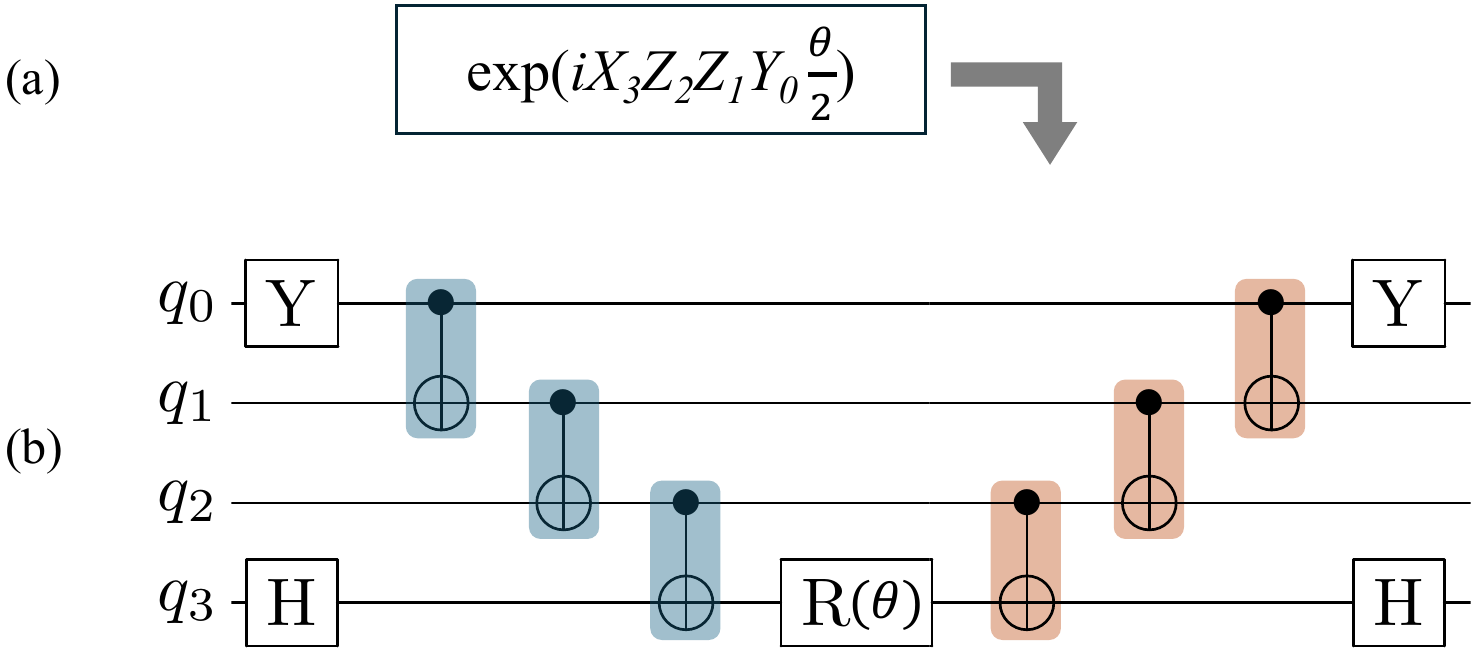}
    \caption{(a) The Pauli string, as a basic element of quantum simulation. (b) The hidden inverses in the circuit for Pauli string. Here we use blue boxes to represent standard $CX$, and orange boxes to represent inverse version $CX^{-1}$.} 
    \label{fig:hamiltonian}
\vspace{-3mm}
\end{figure}

\noindent\textbf{Introducing quantum simulation programs.}
We start with introducing the Pauli string, a basic element in quantum simulation. 
A Pauli string $P$ is made up of Pauli operators $\sigma_i \in \{I,X,Y,Z\}$ on different qubits $q_i$, for example, $P = X_3Z_2Z_1Y_0$. 
As shown in Fig.~\ref{fig:hamiltonian}, the operator $exp(iP\frac{\theta}{2})$ can be compiled into the circuit.
At the beginning and the end of circuit, there are single-qubit gates $H$ and $Y$ on qubits $q_3$ and $q_0$, corresponding to the Pauli operators $X_3$ and $Y_0$. In the middle is a rotation gate, sandwiched by the left $CX$ tree and the right $CX$ tree.
This is the structure of a Pauli string sub-circuit, and the quantum simulation programs are composed of several sub-circuits like this.

\noindent\textbf{Hidden inverses in a quantum simulation circuit.}
Given the Pauli string circuit in Fig.~\ref{fig:hamiltonian}~(b), we can exploit its structure for gate-level optimization. Looking into the left $CX$ tree and right $CX$ tree, they are always symmetric to each other~\cite{li2022paulihedral}. This provides a chance to construct a set of hidden inverses between the left and right $CX$ trees, with each pair of $CX$ are applied on the same \textit{control} and \textit{target} qubits. Here we substitute all the $CX$ in right tree with $CX^{-1}$, using the technique we introduced in Section~IV. 

Besides, our method can also be applied to the pairs of $H$ gates, for they are also self-adjoint unitaries.
By compiling to the $H$ gates on IBM quantum computers, we have $H = R_Z(\frac{\pi}{2})R_X(\frac{\pi}{2})R_Z(\frac{\pi}{2})$. Similarly to $CX^{-1}$, we can have the inverse gate of $H$ as $H^{-1} = R_Z(-\frac{\pi}{2})R_X(-\frac{\pi}{2})R_Z(-\frac{\pi}{2})$.

\noindent\textbf{Cross-layer optimization for quantum simulation circuits.}
By exploiting the structure inside a quantum simulation circuit, we are able to optimize each sub-circuit of a Pauli string, leveraging the gate-pulse co-optimization described in Section~IV. Base on such insight, we build our pipeline for mitigating coherent noise for quantum simulation program:
\begin{enumerate}
    \item When synthesizing the circuit from the high-level program information (e.g. Pauli string sets), we capture the pattern of $CX$ and $H$ "gate pairs", and mark it as a chance for \textit{hidden inverse}. For example, the left and right $CX$ tree for a Pauli string. This information can be integrated into a program-based compiler such as 2QAN~\cite{lao20222qan} and Paulihedral~\cite{li2022paulihedral}.
    
    \item We analyse each gate pairs, and decide each of the gate should be a standard version ($CX$) or inverse version ($CX^{-1}$). This information is attached to each gate for further compilation to pulse, and it can be realized by defining customized gate in IBM Qiskit~\cite{qiskit2024}.

    \item During the control pulse generation, the standard and inverse gates are generated to different pulses, which are constructed using the approach described in Section~IV~A.
\end{enumerate}

\subsection{Cross-Layer Optimization for Universal Quantum Programs}
In the above, we illustrate our program-gate-pulse model for coherent noise mitigation, based on a specific quantum algorithm. Here we want to extend such model to universal quantum programs, emphasizing the generalizability of our optimization model.

\noindent\textbf{A universal quantum computing model.} Here we introduce a quantum gate model by Sleator \& Weinfurter~\cite{sleator1995,barenco1995elementary}, which can compose the arbitrary unitary operations in quantum computing. For any single-qubit unitary $U$, the model give a operator $\wedge_m(U)$, as the multi-control unitary gate. $\wedge_m(U)$ can be regarded as a generalization form of the multi-control Toffoli gates~\cite{miller2011elementary, grosse2009exact}, with $m+1$ input qubits $\ket{x_1,...,x_m,y}$, that applies $U$ to $y$ only when $\wedge_{k=1}^m x_k = 1$. In example, with the Pauli $X$ matrix $\sigma_x$, we have gates $X = \wedge_0(\sigma_x)$, $CX = \wedge_1(\sigma_x)$ and \textit{Toffoli} $ = \wedge_2(\sigma_x)$. Barenco et al.~\cite{barenco1995elementary} provide an approach to decompose arbitrary $\wedge_m(U)$ into basic gates, based on it we elaborate our program-based optimization.

\begin{figure}[h!]
    \centering
    \includegraphics[page=1,width=.47\textwidth]{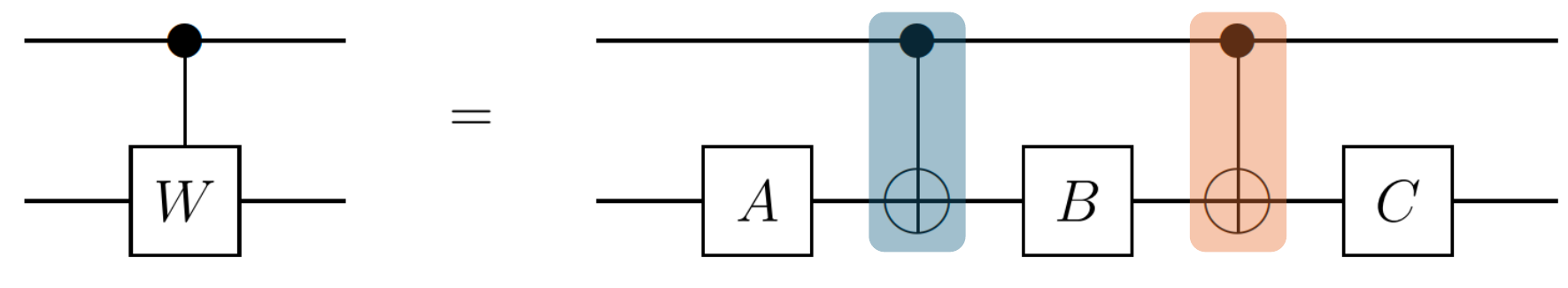}
    \caption{Hidden inverse inside the decomposition of $\wedge_1(U)$ gate.} 
    \label{fig:first_layer}
\end{figure}

\noindent\textbf{Layer-1: Hidden inverse in general $\wedge_1(U)$ gates.} For a single qubit unitary $W$, a $\wedge_1(W)$ gate can be decomposed into the form shown in Fig.~\ref{fig:first_layer}~\cite{barenco1995elementary, miller2011elementary}. There are single qubit unitary $A,B,C$, with $A\cdot B\cdot C = I$ and $A \cdot \sigma_x \cdot B \cdot \sigma_x \cdot C = W$. $A$, $C$ are at the beginning and the end, and $B$ is in the middle, sandwiched by a pair of $CX$ gates. Here we construct the hidden inverse within this pair of $CX$, with the similar approach introduced for $CR_Z$ gate. This is the first layer inside our program-gate co-optimization.

\begin{figure}[h!]
    \centering
    \includegraphics[page=1,width=.47\textwidth]{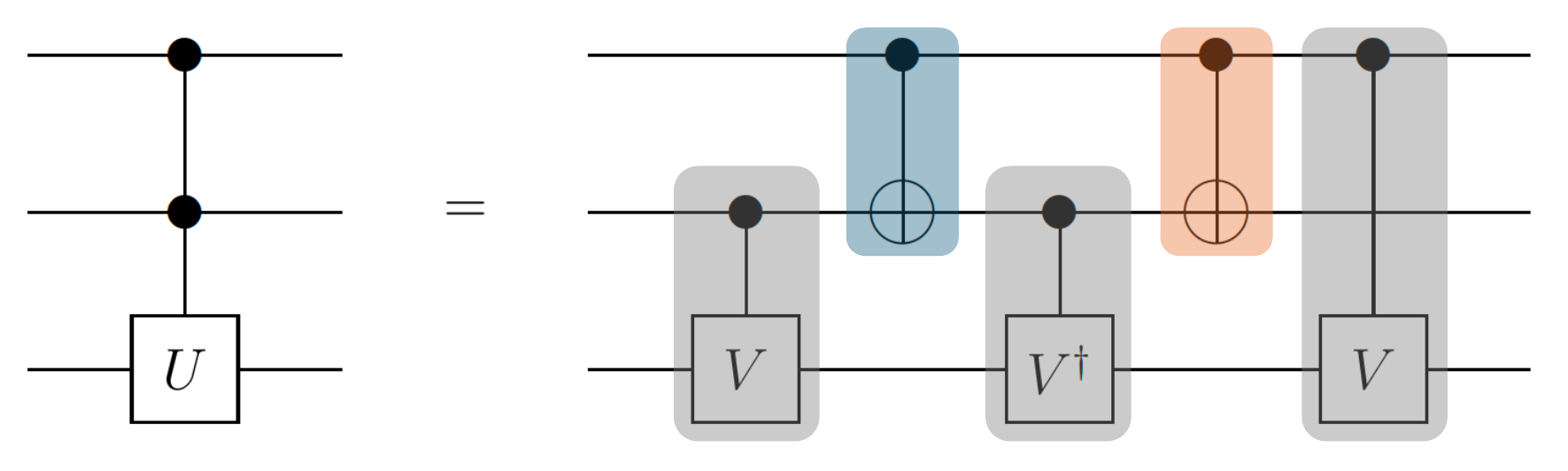}
    \caption{Optimization on the decomposition of $\wedge_2(U)$ gate.} 
    \label{fig:second_layer}
\end{figure}

\noindent\textbf{Layer-2: From $\wedge_2(U)$ gates to $\wedge_1(U)$ gates.}
Based on the $\wedge_1(U)$ situation described previously, we extend the optimization to $\wedge_2(U)$ gates. Shown in Fig.~\ref{fig:second_layer}, a general $\wedge_2(U)$ gate can be decomposed in the form~\cite{barenco1995elementary, miller2011elementary}: a pair of $CX$ sandwiching a $\wedge_1(V^\dagger)$ gate, with other two $\wedge_1(V)$ gates at the beginning and the end. Here we have $V$ also being a single-qubit unitary, that $V^2 = U$. Among these gates, the pair of $CX$ can construct a hidden inverse, while other $\wedge_1(V)$ and $\wedge_1(V^\dagger)$ can be optimized by the layer-1 approach.

\begin{figure}[h!]
    \centering
    \includegraphics[page=1,width=.43\textwidth]{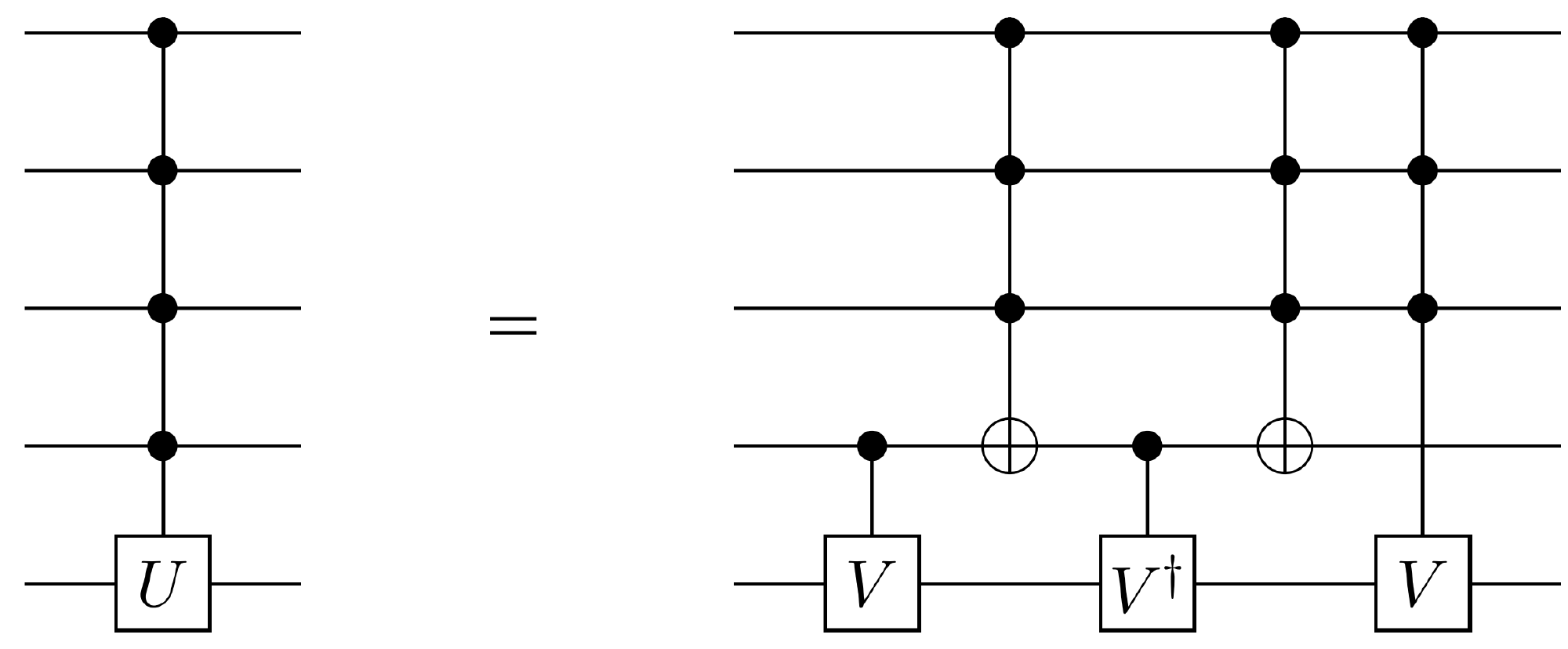}
    \caption{Decomposition of $\wedge_m(U)$ gate.} 
    \label{fig:third_layer}
\end{figure}

\noindent\textbf{Layer-3: Optimizing universal $\wedge_m(U)$ gates.}
Furthermore, we extend out optimization to general $\wedge_m(U)$ gates. In Fig.~\ref{fig:third_layer} is an example of decomposing $\wedge_m(U)$ gate, into two $\wedge_{m-1}(\sigma_x)$, a $\wedge_{m-1}(V)$, $\wedge_1(V)$ and $\wedge_1(V^\dagger)$. Recursively, the $\wedge_{m-1}(U)$ gates can finally be decomposed into a set of $\wedge_2$ and $\wedge_1$ gates. Hence, the optimization problem is simplified to layer-1 and layer-2.

\section{Evaluation}
In this section, we evaluate our cross-layer optimization by running various quantum programs on real quantum computers. Herein we are trying to answer if the proposed cross-layer optimization framework improves the circuit fidelity on a set of major quantum benchmarks, on real quantum hardware.

\subsection{Evaluation Methodology}
\noindent\textbf{Framework Implementation: } 
We extend Qiskit~\cite{qiskit2024} to implement our co-optimization pass. 
We utilize the customized gate feature in Qiskit, to define our $CX^{-1}$ and $H^{-1}$ gates. 
The standard quantum pulse for these gates are retrieved from IBM quantum cloud platform, and modify their parameters to build the pulses of inverse gates.
For the $\wedge_m(U)$ gates and control-rotation gates ($CR_Z, CPhase$), we define the \textit{decompose()} function to integrate our hidden inverse structure.

\noindent\textbf{Experiment Platform: }
Our compiler framework are performed on a Cent-OS 7 server featuring dual Intel Xeon Platinum 8255C CPUs for a total of 48 cores/96 threads at 2.5 GHz, 128 GB of RAM.
Our real hardware execution are performed on IBM Brisbane quantum computer, with Qiskit Pulse~\cite{alexander2020qiskit} support. 







\noindent\textbf{Benchmarks: }
We benchmark and compare our approach with original Qiskit compiler on various algorithms. For the benchmark of quantum simulation programs, we use quantum approximate optimization algorithm for MAX-CUT problems (QAOA)~\cite{farhi2014quantum}, Hamiltonian simulation of three models: Ising model, XY model and Heisenberg model. For other programs, we use Quantum Fourier Transformer Adder (QFT-Adder) and Quantum Phase Estimation (QPE).

\noindent\textbf{Metrics: }
\textit{Fidelity} refers to the similarity between two quantum states. For the probability distributions $p$ and $q$, quantum fidelity is defined as $F(p,q) = (\sum_x\sqrt{p(x)q(x)})^2$. In this work, fidelity measures how similar the results of NISQ devices are to the ideal distribution. Besides, we also calculate the fidelity improvement of our framework compared with the original.

\subsection{Result}
\begin{figure}[t]
    \centering
    \includegraphics[page=1,width=.47\textwidth]{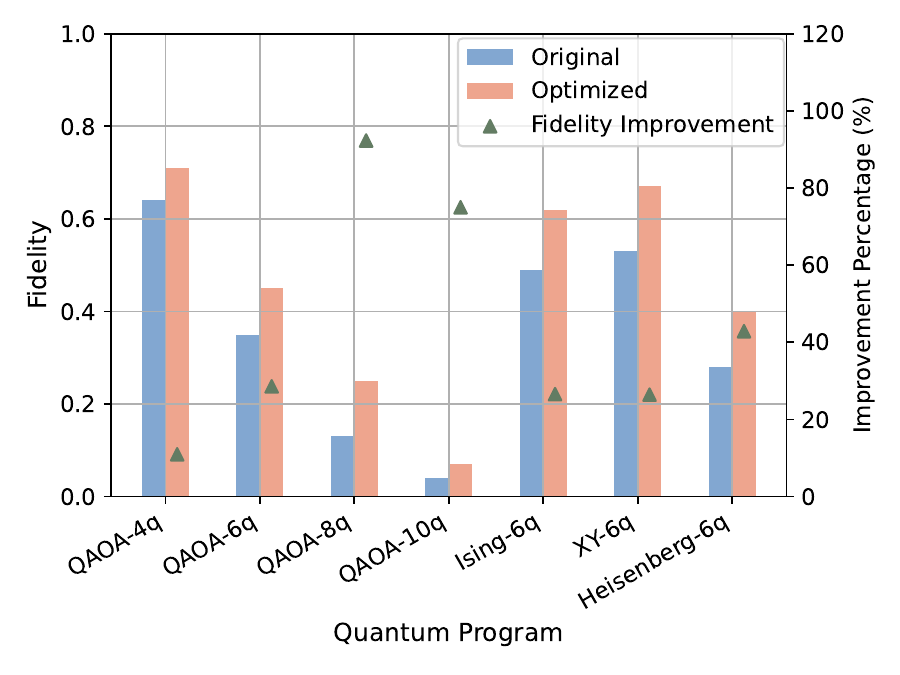}
    \caption{Evaluation on various quantum simulation programs.} 
    \label{fig:res_sim}
\end{figure}

\noindent\textbf{Quantum simulation.}
We choose our quantum Hamiltonian simulation programs from the papers of 2QAN~\cite{lao20222qan} and Paulihedral~\cite{li2022paulihedral}. Here we have four different sizes of circuits for QAOA -- 4-qubit, 6-qubit, 8-qubit and 10-qubit, also 6-qubit circuits for these three Hamiltonian simulation models. As the result we provide in Fig.~\ref{fig:res_sim}, our framework shown fidelity improvement to varying degrees, comparing with the original compiler. The percentage of fidelity improvement are 11\%, 29\%, 92\%, 75\%, 27\%, 26\% and 43\% respectively, with an average of 43\% improvement.

\noindent\textbf{QFT-Adder and QPE.}
For the case of universal quantum programs, we choose QFT-Adder and QPE as the benchmarks. Here we generate 5-qubit, 7-qubit and 9-qubit circuits for QFT-Adder, and 5-qubit, 6-qubit and 7-qubit circuits for QPE. Compared with the original execution result, our framework improve the fidelity by 23\%, 57\%, 67\%, 36\%, 47\% and 60\% respectively, with an average of 48\% improvement.

\begin{figure}[t]
    \centering
    \includegraphics[page=1,width=.47\textwidth]{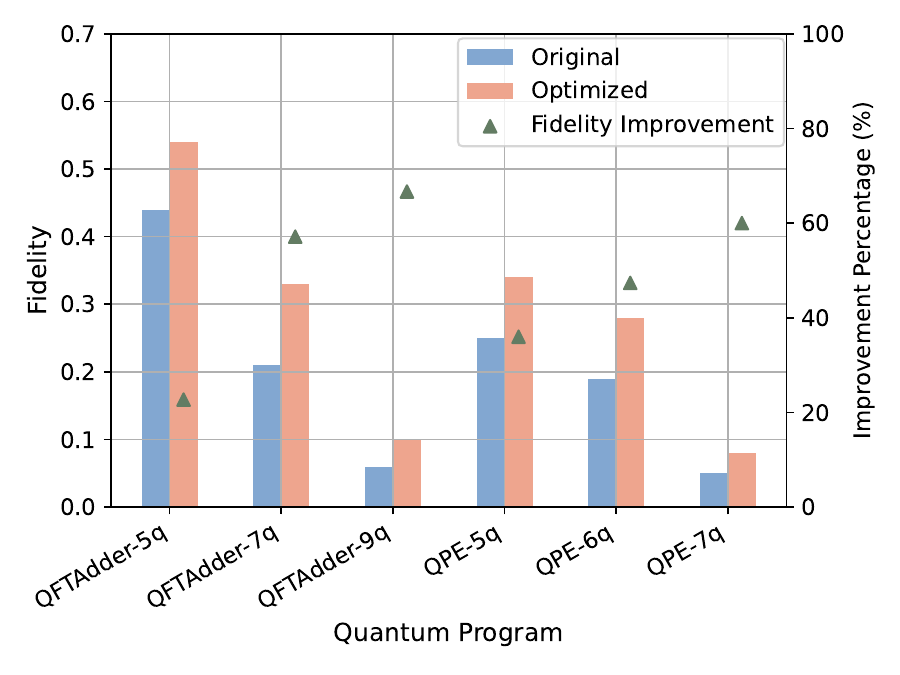}
    \caption{Evaluation on QFT and QPE quantum algorithms.} 
    \label{fig:res_other}
\vspace{-3mm}
\end{figure}

\section{Conclusion}
This paper demonstrates that it is possible to reduce coherent error on quantum computers by cross-layer optimizations, including pulse-level, gate-level and program-level.
The quantum computer we used for the paper is the IBM Brisbane, on which we ran a series of different benchmarks including QAOA, Hamiltonian simulation, QFT-Adder and QPE, achieving a maximum 92\% improvement of circuit fidelity, with average 45\% improvement. 
While we demonstrated our technique on a superconducting quantum computer, our framework is also adaptable to other quantum computing hardware, like trapped-ion, by modifying the pulse-level method~\cite{majumder2023characterizing}.

\clearpage
\bibliographystyle{plain}
\bibliography{references}

\end{document}